\documentclass[12pt]{article}

\usepackage{lscape}
\usepackage{dcolumn}
\newcolumntype{d}[1]{D{.}{.}{#1}}
\usepackage{epsfig}
\usepackage{amsmath}
\usepackage{hhline}
\usepackage{amssymb}
\usepackage{times}
\usepackage{cite}

\newlength{\dinwidth}
\newlength{\dinmargin}
\begin{document}

\newcommand{\F}{$ F_{2}(x,Q^2)\:$} 
\newcommand{\FL}{$ F_{L}(x,Q^2)\:$}
\newcommand{\FLD}{$ F_{L}^{D}(x_{IP},Q^2,\beta)\:$}
\newcommand{\Fc}{$ F_{2}\,$}
\newcommand{\Q}{$ Q^{2}\,$}
\newcommand{\gv}{\,GeV$^2\,$}
\newcommand{\as}{$\alpha_s\,$}
\newcommand{\amz}{$\alpha_s(M_Z^2)\,$} 
\newcommand{\bs}{\overline{s}}
\newcommand{\bu}{\overline{u}}
\newcommand{\bd}{\overline{d}}
\newcommand{\bq}{$ \overline{q}$} 
\newcommand{\xbq}{$x \overline{q}$}    
\newcommand{\FLc}{$ F_{L}\,$} 
\newcommand{\FLDc}{$ F_{L}^{D}\,$} 
\newcommand{\xg}{$xg(x,Q^2)\,$}
\newcommand{\xgc}{$xg\,$}
\newcommand{\qc}{$q\,$}
\newcommand{\ipb}{pb$^{-1}\,$} 
\newcommand{\pdff}{$\partial F_{2} / \partial \ln Q^{2}\,$ }
%
%
%%%%%%%%%%%%%%%%%%%%%%%%%%%%%%%%%%%%%%%%%%%%%%%%%%%%%%%%%%%%%%%%%%%%%%%%%%%%%%%

\begin{titlepage}

\noindent
\begin{flushright}
DESY 08-053\hfill ISSN 0418-9833\\
May 2008
\end{flushright}
\vspace*{2cm}

\begin{center}
\begin{Large}

{\bf 
Measurement of the Proton Structure Function \\
{\boldmath $F_L(x,Q^2)$} at Low {\boldmath $x$}
 \\ 
}

\vspace{2cm}

H1 Collaboration

\end{Large}
\end{center}

\vspace{2cm}

\begin{abstract}
\noindent
   A first measurement is reported of the longitudinal proton
   structure function  $F_L(x,Q^2)$ 
   at the $ep$ collider HERA. It is based on inclusive deep inelastic
   $e^+p$ scattering cross section measurements
   with a  positron beam  energy of $27.5$\, GeV  and proton beam
   energies  of $920$, $575$ and $460$\,GeV.
   Employing the energy dependence of the cross section, $F_L$
   is measured in a range of  squared 
   four-momentum transfers  $12 \leq Q^2 \leq 90~$GeV$^2$ and 
   low  Bjorken $x$ $0.00024 \leq x  \leq  0.0036$.
   The  \FLc values agree with higher order QCD calculations
   based on parton densities obtained using  
   cross section data previously measured at HERA.
\end{abstract}

\vspace{0.5cm}

\begin{center}
Submitted to Phys. Lett. {\bf B}
\end{center}

\end{titlepage}

%%%%%%%%%%%%%%%%%%%%%%%%%%%%%%%%%%%%%%%%%%%%%%%%%%%%%%%%%%%%%%%%%%%%%%%%%%%%%%%

\begin{flushleft}
%-- H1AUTS Author list by names 
%-- Status: Wed Apr  2 09:26:46 CEST 2008  Number of authors = 273 

F.D.~Aaron$^{5,49}$,           %BUCH-PD        11/06           Aaron               
C.~Alexa$^{5}$,                %BUCH-PD        06/06           Alexa               
V.~Andreev$^{25}$,             %LPI -PD        8/88            Andreev             
B.~Antunovic$^{11}$,           %DESY-PD        05/07           Antunovic           
S.~Aplin$^{11}$,               %DESY-LEFT      01/08           Aplin               
A.~Asmone$^{33}$,              %ROME-ST        07/2            Asmone              
A.~Astvatsatourov$^{4}$,       %BRUX-LEFT      01/08           Astvatsatourov      
A.~Bacchetta$^{11}$,           %DESY-LEFT      02/08           Bacchetta           
S.~Backovic$^{30}$,            %PODG-PD        03/2            Backovic            
A.~Baghdasaryan$^{38}$
% ...

%-- H1 Institutes 
\bigskip{\it
% ...
$ ^{2}$ Vinca  Institute of Nuclear Sciences, Belgrade, Serbia \\
% ...
$ ^{4}$ Inter-University Institute for High Energies ULB-VUB, Brussels;
          Universiteit Antwerpen, Antwerpen; Belgium$^{ c}$ \\
 $ ^{5}$ National Institute for Physics and Nuclear Engineering (NIPNE) ,
          Bucharest, Romania \\
% ...
$ ^{11}$ DESY, Hamburg, Germany \\
% ...
$ ^{25}$ Lebedev Physical Institute, Moscow, Russia$^{ e}$ \\
% ...
$ ^{30}$ Faculty of Science, University of Montenegro,
          Podgorica, Montenegro$^{ e}$ \\
% ...
$ ^{33}$ Dipartimento di Fisica Universit\`a di Roma Tre
          and INFN Roma~3, Roma, Italy \\
% ...
$ ^{38}$ Yerevan Physics Institute, Yerevan, Armenia \\
% ...
$ ^{49}$ Also at Faculty of Physics, University of Bucharest,
          Bucharest, Romania \\
% ...
$ ^c$ Supported by FNRS-FWO-Vlaanderen, IISN-IIKW and IWT
      and  by Interuniversity
Attraction Poles Programme,
      Belgian Science Policy \\
% ...
$ ^e$ Supported by the Deutsche Forschungsgemeinschaft \\
% ...
}
\end{flushleft}
\newpage
%%%%%%%%%%%%%%%%%%%%%%%%%%%%%%%%%%%%%%%%%%%%%%%%%%%%%%%%%%%%%%%%%%%%%%%%%%%%%%%
%\linenumbers
\section{Introduction}
This letter presents the first measurement 
of the longitudinal structure function $F_L(x,Q^2)$ of the proton
at low Bjorken $x$. The inclusive deep inelastic $ep$ scattering (DIS)
cross section at  low $Q^2$, written in reduced form as
\begin{equation}
 \sigma_r(x,Q^2,y) = \frac{d^2\sigma}{dxdQ^2} 
 \cdot \frac{Q^4 x}{2\pi \alpha^2 Y_+}  
  =  F_2(x,Q^2) - \frac{y^2}{Y_+} \cdot  F_L(x,Q^2)~,
       \label{sig}
  \end{equation}  
is determined by two structure functions,  \Fc and $F_L$.
Here, $Q^2 = -q^2 $ is the negative four-momentum squared
transferred between the electron\footnote{The term electron is
used here to denote both electrons and positrons unless the charge state
is specified explicitely. The data analysed are from positron-proton
scattering, except for some measurements of background properties
which additionally include electron-proton scattering data.}
 and the proton, and $x=Q^2/2qP$ denotes
the Bjorken variable, where $P$ is the four-momentum of the proton. 
The two variables are related through the inelasticity  of
the scattering process, $y=Q^2/sx$,
where $s=4E_eE_p$ is the centre-of-mass energy
squared determined  from the electron and proton beam 
energies, $E_e$ and $E_p$. In equation~\ref{sig},
$\alpha$ denotes the fine structure constant and $Y_+=1+(1-y)^2$.
\par

The two proton structure functions \FLc and \Fc are of complementary 
nature.  They are related to the $\gamma^* p$ interaction cross sections
of longitudinally and transversely polarised virtual photons,
$\sigma_L$ and $\sigma_T$, according to
$F_L \propto \sigma_L$  and $F_2 \propto (\sigma_L + \sigma_T)$.
Therefore the relation $0 \leq F_L \leq F_2$  holds. 
 In the Quark Parton Model (QPM),
 \Fc is the sum of the quark and anti-quark $x$ distributions,
weighted by the square of the electric quark charges, whereas the value of \FLc is
zero~\cite{cgross}. In Quantum Chromodynamics (QCD), 
the longitudinal structure function differs from zero, receiving
contributions from quarks and from gluons\,\cite{am}. At low $x$
and in the $Q^2$ region of deep inelastic scattering the
gluon contribution greatly exceeds the quark contribution.
%part due to  gluons exceeds the part due to  quarks by far. 
Therefore
\FLc is a direct measure of the gluon distribution
to a very good approximation. 
The gluon distribution is also constrained by the scaling 
violations of \F as described by the DGLAP QCD evolution 
equations\,\cite{dglap}.  An independent measurement of \FLc at HERA,
and its comparison with predictions derived from the
gluon distribution extracted from the $Q^2$ evolution of
$F_2(x,Q^2)$, thus represents a crucial test 
on the validity of perturbative QCD  at low Bjorken $x$.

\par

The longitudinal structure function, or the equivalent cross section
ratio $R=\sigma_L/\sigma_T=F_L/(F_2-F_L)$, was measured
previously in fixed target experiments ~\cite{rdata} and 
found to be small  at 
large $x \geq 0.2$,  confirming the QPM prediction
in the $Q^2$ region of DIS.
%in the deep inelastic $Q^2$ region.
\par

From experimental determinations by H1\,\cite{fla,alfs,h1pdf2k},
which used assumptions on the behaviour of \Fc
in extracting \FLc, and from theoretical analyses of the
inclusive DIS cross section data \cite{mrsw,cteq},
the longitudinal structure function at low $x$ is
expected to be significantly larger than zero.
This prediction relies on perturbative QCD calculations
of \FLc to next-to-leading order (NLO)\,\cite{willy} and
NNLO\,\cite{jos}.
%From theoretical analyses of the inclusive DIS cross section data \cite{mrsw,cteq}
%and from experimental determinations by H1\,\cite{fla,alfs,h1pdf2k},
%which used assumptions on the behaviour of \Fc
%in extracting \FLc, the longitudinal
%structure function at low $x$ is expected to be significantly larger than zero.
\par

The measurement of \FLc  requires several sets of  DIS cross sections
at fixed $x$ and $Q^2$ but at different $y$. This was 
achieved at HERA by variations of the proton beam energy 
whilst keeping the lepton beam energy fixed. 
The sensitivity to \FLc is largest at high $y$ as 
its contribution to $\sigma_r$ is proportional to $y^2$.
At low  $Q^2$, high $y$ values correspond to
low values of the scattered electron energy.
Small energy depositions can also be caused by hadronic final state 
particles leading to fake electron signals.
These are dominantly due to photoproduction processes at $Q^2 \simeq 0$.  
The large size of this background makes the measurement of \FL
particularly challenging.

The present measurement of \FL is based on  data
collected with the H1 detector in
$e^+p$ collisions from January to June 2007
with a positron beam energy  of $27.5$\,GeV.
Three proton beam energies were used, the largest, nominal
energy of $920$\,GeV, the smallest energy
of $460$\,GeV and an intermediate energy of $575$\,GeV, chosen for
an approximately equal span between the three resulting cross
section measurements in $y^2/Y_+$ (see equation\,\ref{sig}). The integrated
luminosities collected with H1 are $21.6$\,pb$^{-1}$, $12.4$\,pb$^{-1}$ and 
$6.2$\,pb$^{-1}$, respectively.  This letter presents first results on \FLc  
in an intermediate range of $Q^2$, between 12 and 90\,GeV$^2$.
\section{Data Analysis}
\subsection{H1 Detector}
The H1 detector\,\cite{h1} was built and upgraded
for the accurate measurement of 
%inelastic
$ep$ interactions  at HERA.  The detector components
most relevant to this measurement 
are the central jet drift chamber (CJC),
the central inner proportional chamber (CIP), the backward lead-scintillator
calorimeter (SpaCal) and the liquid argon calorimeter (LAr).
The CJC measures transverse momenta of tracks with an accuracy of 
$\delta p_t/p_t^2 \simeq 0.005$/GeV.  
Complementary tracking information is obtained from  the
backward silicon tracker (BST), which is positioned around the
beam pipe, and from the $z$ drift chamber COZ,
which is located in between the two cylinders of the CJC.
The CIP provides  trigger information on central tracks\,\cite{cip}.
The SpaCal\,\cite{spacal} has an energy resolution
of $\delta E/E \simeq 0.07/\sqrt{E/\mathrm{GeV}}$ for electromagnetic
energy depositions and is complemented by a hadronic 
section. It also provides a trigger down to $2$\,GeV energy. The LAr 
allows the hadronic final state to be reconstructed
with an energy resolution of about $0.50/\sqrt{E/\mathrm{GeV}}$. 

Photoproduction events can be tagged with an electron calorimeter placed
at $z=-6$\,m downstream in the electron beam direction, which defines the
negative $z$ axis and thus the backward direction. The luminosity
is determined from the Bethe-Heitler scattering process, which
is measured using a photon calorimeter at $z=-103$\,m.
\subsection{Kinematic Reconstruction and Event Selection}
The DIS kinematics at large $y$ are most accurately
reconstructed using the polar angle, $\theta_e$,
and the energy, $E_e'$, of the scattered electron according to
\begin{equation}
y = 1 - {E_e' \over E_e} {\sin}^2 (\theta_e /2)~, \; \; \; \; \;
\; \;
{Q}^2 = {{E_e'}^2 {\sin} ^ 2 \theta_e \over 1-y}~, \enspace
\end{equation}
where $x=Q^2/sy$. The event signature of this analysis comprises an electron
scattered backwards and a well reconstructed
event vertex. The scattered electron energy is measured in the 
backward calorimeter 
SpaCal. The polar angle is determined by the positions of the
interaction vertex and  the electron cluster in the SpaCal.

In order to trigger on low energy depositions with a 
threshold of $2$\,GeV, a dedicated trigger was developed based on the SpaCal
cell energy depositions.  At small energies the SpaCal trigger
is complemented  by the CIP track trigger which reduces the
trigger rate to an acceptable level. 
The efficiency of this high $y$  trigger 
is constant at around $98$\,\% above $3$\,GeV, 
as monitored with independent triggers. At energies larger 
than $7$\,GeV no track condition is used in the trigger and the
efficiency, up to highest energies, exceeds $99$\,\%.

The event selection is based on the identification of the scattered electron as a localised energy deposition (cluster) of more than $3.4$\,GeV in the SpaCal.
Hadrons, dominantly from photoproduction but also from DIS, may also
lead to such energy depositions. This fake electron
background is  reduced by the requirement of a
small transverse size of the cluster, $R_{log}$, which is
estimated using a logarithmic energy weighted 
cluster radius. The background is further reduced
by the requirement that the energy behind the cluster,
measured in the hadronic part of the SpaCal, may not
exceed a certain fraction of $E_e'$. For lower
energies the selected  cluster must be linked to a track.
If the highest energy cluster fails to fulfill the
selection criteria, the next to highest energy cluster
passing the selection criteria is considered. Alternatively ordering the SpaCal clusters according to the scattering angle or transverse momentum
gives consistent cross section results.

An additional suppression of   photoproduction 
background is achieved by requiring  longitudinal energy-momentum 
conservation using the variable 
\begin{equation}
E - p_z = \Sigma_i (E_i - p_{z,i}) + E_e'(1 - \cos \theta_e),  
%\enspace
\label{empz}
\end{equation}
which for genuine, non-radiative DIS events is approximately equal to 2$E_e$.
Here $E_i$ and $p_{z,i}$ are the energy and longitudinal momentum
component of a particle $i$ in the hadronic final state.
This requirement also suppresses events with hard initial
state photon radiation. QED Compton events are excluded using a topological
cut against two back-to-back energy depositions in the SpaCal.

\begin{table}[tb]
\begin{center}
\begin{tabular}{|l|l|} \hline
 Energy $E_e'$ of scattered electron candidate & $> 3.4$~GeV  \\ 
 Transverse size $R_{log}$ of candidate cluster  & $< 5$~cm \\ 
 Hadronic energy fraction behind the cluster  & $< 15\,\%$ of ${E_e}'$ \\
 Transverse distance between cluster and linked track   & $< 6$~cm \\
  $E-p_z$  & $> 35$~GeV \\ 
 $z$ position of interaction vertex & $|z_v| < 35$~cm \\ \hline
\end{tabular}
\caption{Criteria applied to select DIS events at high inelasticity $y$. }
\label{tab:tabcuts}
\end{center}
\end{table}

The selection is optimised to obtain large detection efficiency. 
This required detailed studies which were also  based   
on high statistics event samples  obtained in the years 2003-2006,
corresponding to $51$\,pb$^{-1}$ of e$^+$p 
and $45$\,pb$^{-1}$ of e$^-$p interactions
taken with a dedicated high $y$ trigger at $920$\,GeV
proton beam energy.  The event selection criteria  for the high $y$ region
are summarised in table~\ref{tab:tabcuts}.

The extraction of 
\FLc also requires  the measurement of  cross sections at lower $y$.
The low $y$ region is defined for the
$460$ and $575$\,GeV data with $y < 0.38$ and
for the $920$\,GeV data with $y < 0.5$. The analysis uses a method based on the
electron variables for reconstruction and hence is limited 
to $y \geq 0.1$ for all data sets.
The data at low $y$ involve large 
polar angles $\theta_e$ outside the acceptance of the CJC.
Therefore in this kinematic  region no
link to CJC  tracks is required.  At low $y$ the photoproduction
background is small and  further reduced  by a tightened
cut on $R_{log}<4$\,cm.
\subsection{Background Identification and Subtraction}
At low $E_e'$, corresponding to high $y$,
the remaining background contribution after the event selection may be of a size comparable to or even exceeding the genuine DIS signal. 
The method of background subtraction
relies on the determination of the  electric charge of the electron 
candidate from the curvature of the associated track. 

\begin{figure}[h]
   \centering
    \epsfig{file=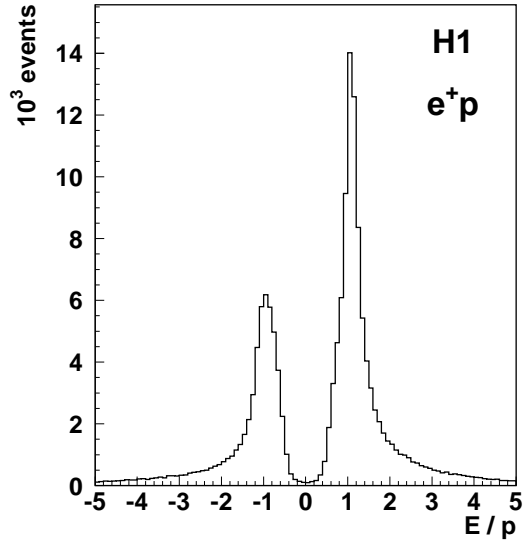,width=8.cm}
   \caption{Distribution of energy  over momentum  for tracks 
    linked to clusters in the SpaCal
    with energy from $3.4$ to $10$\,GeV that pass all the cuts listed in table\,1.
    Tracks with a negative charge are assigned a negative $E/p$.}
   \label{Fig:eoverp}
\end{figure}

Figure\,\ref{Fig:eoverp} shows the 
$E/p$ distribution of the scattered electron candidates  
from $e^+p$ interactions
with the energy $E$ measured in the SpaCal
and the  momentum $p$ of the linked track determined by the CJC.
The good momentum resolution leads to a clear distinction between the
negative and positive charge distributions.   
The smaller peak corresponds to tracks with negative charge 
and thus represents almost pure background. These tracks are termed 
wrong sign tracks.   The higher peak, due to right sign tracks,
contains the genuine DIS signal superimposed on the  
remaining positive  background. 
The size of the latter to first
approximation equals the wrong sign background.  The principal
method of background subtraction, and thus of measuring the
DIS cross section up to $y \simeq 0.9$, consists of the
subtraction of the wrong sign from the right sign event distribution
in each $x,Q^2$ interval. 

The background subtraction based on the charge measurement
requires a correction for a small but non-negligible charge asymmetry
in the  negative and positive background samples,
as has been observed previously by H1\cite{alfs}.
The main cause for this asymmetry lies in the  enhanced
energy deposited by anti-protons compared to protons at
low energies. The most precise measurement of the  background charge 
asymmetry  has been obtained from comparisons of samples
of negative tracks in $e^+p$ scattering with samples of positive 
tracks in $e^-p$ scattering.  An asymmetry ratio of  negative
to positive tracks of $1.057 \pm 0.006$ is measured using the high statistics
$e^{\pm}p$ data collected by H1 in 2003-2006. This result is verified using 
photoproduction events, with a tagged scattered
electron, for which an asymmetry ratio
of $1.06 \pm 0.01$ is measured. The difference in the hadronic final
state between low and high proton beam energy data samples
leads to an additional uncertainty of $0.003$ on the asymmetry ratio.

The photoproduction background to the $E_{p}=920$\,GeV data, which
are analysed at lower $y$ than the low $E_p$  data, is subtracted using 
a PHOJET~\cite{phojet} simulation normalised to the tagged photoproduction data.
This background estimate agrees well with the corresponding
result from the wrong sign analysis at high $y$.
\subsection{Comparison of Data with Simulations}
\begin{figure}[t]
   \centering
  \epsfig{file=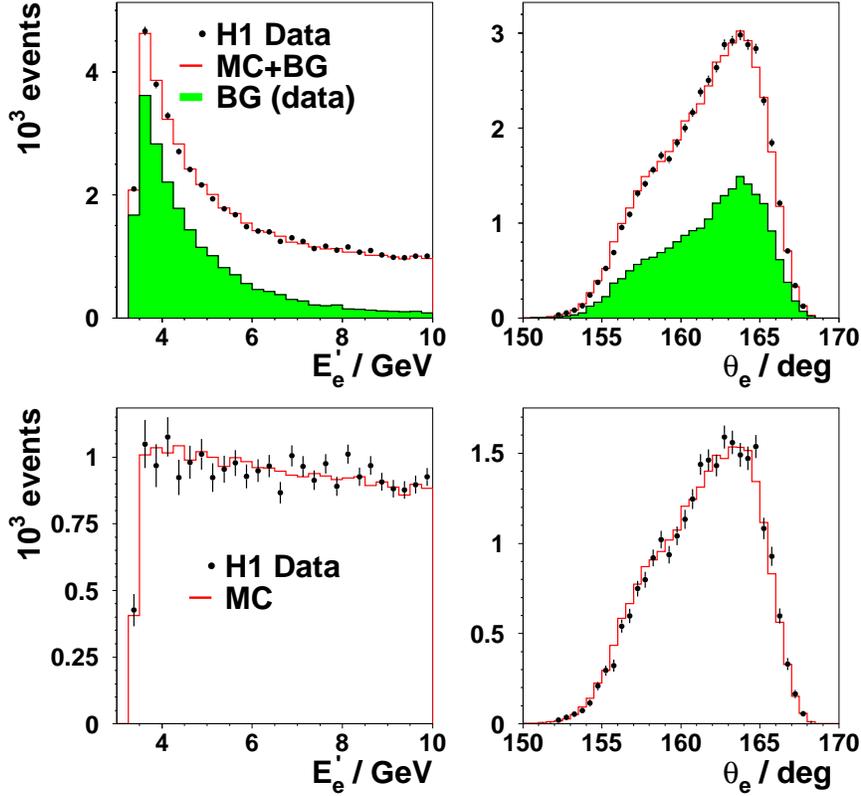,width=12cm}
      \caption{Top: comparison of the correct sign data
(points) with the sum (open histogram) of the DIS MC
simulation and background, determined from
the wrong sign data (shadowed histogram),
for the energy $E_e'$ (left) and the
polar angle $\theta_e$ (right) of the scattered electron,
for the $460$\,GeV data with $E_e' < 10$\,GeV. Bottom: as top but
after background subtraction.
}
\label{Fig:highy460}
\end{figure}
High statistics  Monte Carlo (MC)
simulations of DIS events are performed for the three proton beam energies using the 
DJANGO program~\cite{django}, which includes
leading order QED radiative corrections. The hadronic 
final state is simulated using ARIADNE~\cite{ariadne}, based on
the Color Dipole Model, with subsequent fragmentation
as described in JETSET~\cite{jetset}.  The detector response is simulated
using a program based on GEANT\,\cite{geant}. The simulated events are
subject to the same reconstruction and analysis software
as the data.  
The MC simulation uses a QCD
parameterisation of the structure functions~\cite{h1pdf2k} normalised to the measured cross section.

Figure~\ref{Fig:highy460}  shows, as an example,  comparisons 
of the $460$\,GeV high $y$  data with simulated 
distributions, for the  energy and the polar angle
of the scattered electron prior to  and after 
subtraction of the background  
which  is determined using wrong sign data events.
The DIS MC simulation corresponds to correct
sign events with a small contribution from the wrong sign
events subtracted. The latter are caused by events
from lower $Q^2$ which can mimic an electron cluster at larger 
$Q^2$ and also by charge misidentification for the DIS 
events at the appropriate $Q^2$.
The electron energy distribution after
background correction is almost uniform.
A similarly good agreement of the simulation with data 
has been observed for all other physics and technical variable
distributions of relevance to this analysis, for all three
data sets considered.
\section{Cross Section Measurement}
The scattering cross section is measured  in the 
range   $12 \leq Q^2\leq 90$\,GeV$^2$  for
Bjorken $x$ of  $0.00024\leq x \leq 0.015$. The longitudinal
structure function \FL is extracted from three measurements of
$\sigma_r$ at fixed $(x,Q^2)$ but different $y=Q^2/sx$. The data
at lower $E_p$ cover the higher $y$ region. In the present analysis
the cross section measurement is restricted to 
$0.1\leq y \leq 0.56$ at $E_p=920$\,GeV and to $0.1\leq y \leq 0.9$
at  $460$ and $575$\,GeV.

The measurement of  \FLc as described below relies on an
accurate determination of the variation of the cross section for a given $x$ and $Q^2$
at different beam energies. In order to reduce the uncertainty related to the luminosity measurement, which presently is known to 5\% for each proton beam energy of the 2007 data used here,
the three data samples are   
normalised relatively to each other. The renormalisation
factors are determined at low $y$, where the cross section is determined 
by \F only, apart from a small correction due to $R$.
Using weighted  means of cross section ratios,
extended  over bins at low $y$,  relative normalisation factors
are derived to be $0.980$, $0.995$ and  $1.010$ for the
$920$, $575$ and $460$\,GeV data, respectively.  The relative 
normalisation is known to within $1.6$\%. This uncertainty 
comprises a systematic error of $1.4$\%, a statistical error of
$0.6$\% and the residual influence of $R$ is estimated to be $0.3$\%. 

\begin{figure}[h]
   \centering
      \epsfig{file=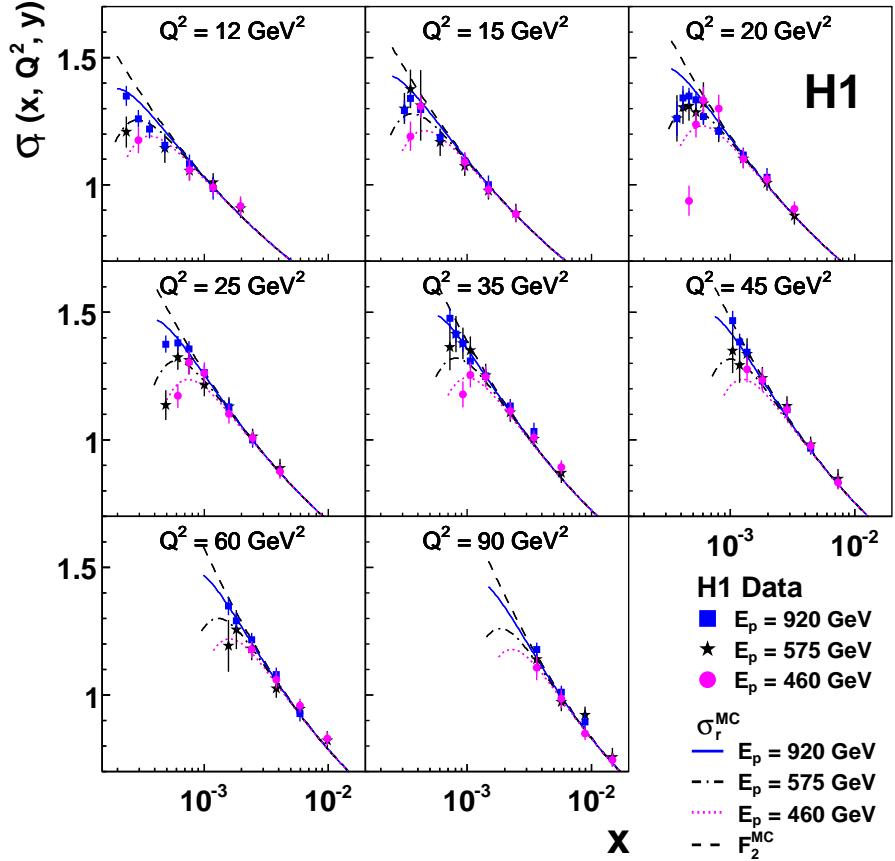,width=14.cm} 
   \caption{The reduced inclusive DIS cross sections
measured at different $Q^{2}$ values and shown as a function of $x$
for the data taken at the three proton beam energies,  $920$\,GeV
(squares), $575$\,GeV (stars) and $460$\,GeV (points).
The error bars represent the statistical and systematic errors added in quadrature.
The absolute luminosity uncertainty of the cross section
measurement is not included in the error bars.
Curves  for $\sigma_r$  as implemented in the
Monte Carlo simulation of the data are shown as  solid  ($920$\,GeV), 
dashed-dotted ($575$\,GeV) and dotted lines ($460$\,GeV)
while the dashed line represents $F_2(x,Q^2)$, which is independent of~$s$. 
}
   \label{figredxsec}
\end{figure}

After background subtraction the data are corrected for
detector efficiencies and for acceptances 
using the Monte Carlo simulations.
The measured differential cross sections are  consistent with the
previous H1 measurement\,\cite{alfs}. They are shown in 
figure\,\ref{figredxsec}.  At large $x$ values 
$\sigma_{r}\approx F_{2}$ and the three measurements
are in good agreement.  The cross sections  rise towards low $x$ 
but are observed to flatten and eventually  turn over at very low 
$x$,  corresponding to high values of $y$,
where \FLc is expected to contribute. This behaviour
is consistent with the expectation as is illustrated
using the cross section as implemented in the
Monte Carlo simulation of the data. 

%%%%%%%%%%%%%%%%%%%%%%%%%%%%%%%%%%%%%%%%%%%%%%%%
%%%%%%%%%%%%%%%%%%%%%%%%%%%%%%%%%%%%%%%%%%%%%%%%%%
%
The systematic uncertainty on the cross section 
is derived from various contributions, some of which  depend
on the $y$ region. The uncertainties leading
to kinematic correlations are:
\begin{itemize}
\item{The uncertainty on the SpaCal electromagnetic energy scale,
determined with the double-angle method,
is $0.4\%$ at large energies  degrading to $1$\% at $3$\,GeV energy.  
This is verified at the kinematic peak, where $E_e'$ has to be
close to $E_e$,
and at lower energies with $\pi^0 \rightarrow \gamma \gamma$,
$J/\Psi \rightarrow e^+ e^-$ decays and with elastic QED Compton events.}
\item{The uncertainty on the electron polar angle is $1$\,mrad, estimated
using independent track information from the BST, the COZ and the CJC.}
\item{The hadronic energy scale, calibrated using 
electron-hadron transverse and longitudinal momentum balance,
has an uncertainty of $4\%$.}
\item{The background charge asymmetry is known to $0.6$\%
based on studies of wrong charge data in $e^{\pm}p$ scattering
and tagged photoproduction events.}
\item{The normalisation of the PHOJET simulation, used for
background subtraction in the $920$\,GeV  data, 
has a $30$\% uncertainty.}
\item{The central track-cluster link efficiency is
verified with an independent track reconstruction using  
BST and  CJC hit information. The uncertainty of this
link efficiency combined with the interaction vertex reconstruction
efficiency is estimated to be $1.5\%$. At low $y$, 
where no track link is required, the remaining
uncertainty from the vertex reconstruction is $0.5$\%.}
\end{itemize}
The uncorrelated systematic uncertainties originate from
the Monte Carlo statistical errors and from the following sources:
\begin{itemize}
\item{The uncertainty on the charge measurement
is determined from data to Monte Carlo comparisons at low $y$ 
and cross checked with radiative events  which are background free 
in the low energy region. As
the charge misidentification causes signal events to be subtracted
as background,  a $1$\% uncertainty on $\sigma_r$ is obtained.}
\item{The radiative corrections are efficiently reduced to below $10$\,\%
by the $E-p_z$ constraint and the topological cut against
QED Compton events. A comparison of calculations based
on the Monte Carlo simulation with the numerical program
HECTOR\,\cite{HECTOR} results in an  uncertainty 
on $\sigma_r$ of $1$\% at high $y$ and $0.5$\% at low $y$.}
\item{The trigger efficiency, determined from independent
monitor triggers, is known to within $1\%$ for the
combined CIP--SpaCal trigger and
$0.5\%$ for the inclusive SpaCal trigger.}
\item{
Comparisons between different electron identification algorithms
and between data and simulations yield
an estimated  uncertainty of $1$\% ($0.5$\%) on the electron identification
at high (low) $y$ in the SpaCal calorimeter.}
\end{itemize}
Further uncertainties, such as the 
effect of the LAr noise on the cross section, have been 
investigated and are found to be negligible.
\begin{figure}[t]
   \centering
   \epsfig{file=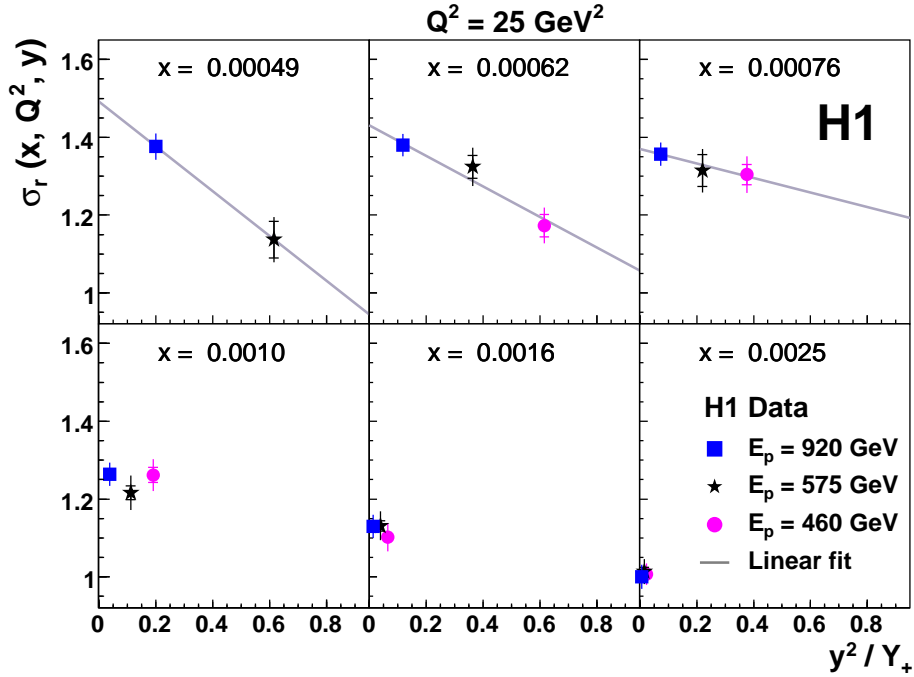,width=13cm} % 
   \caption{The reduced inclusive DIS cross section
 plotted as a function of $y^2/Y_+$
    for six values of $x$  at $Q^2=25$\,GeV$^2$,
    measured for proton beam energies of $920$, $575$ and $460$\,GeV.
    The inner error bars denote the statistical error, the full
    error bars include the systematic errors. The luminosity uncertainty
    is not included in the error bars.
     For the first three bins in $x$, corresponding to larger~$y$, 
    a straight line fit is shown, the slope of which determines $F_L(x,Q^2)$.}
   \label{figros}
\end{figure}
The subtraction of background using wrong sign tracks causes an additional
statistical uncertainty which is included in the statistical error.  
The  correlated and  uncorrelated systematic errors combined with
the statistical error lead to an uncertainty
on the measured cross sections at high $y$  of  $3$ to $5$\%, 
excluding the common luminosity error.
\section{Measurement of  {\boldmath $F_L(x,Q^2)$}}
The longitudinal structure function is extracted from the measurements
of the reduced cross section as the slope of
$\sigma_r$ versus $y^2/Y_+$, as can be seen in equation\,\ref{sig}. 
This procedure is illustrated in figure\,\ref{figros}.  
At a given $Q^2$ value, the lowest $x$ values are generally
accessed by combining only the $920$ and the $575$\,GeV data.
At larger $x$, cross section measurements from all three
data sets are available. These measurements are observed to be
consistent with the expected linear dependence.

The central \FLc values are determined in 
straight-line fits to $\sigma_r(x,Q^2,y)$ as a function of
$y^2/Y_+$ using
the statistical and uncorrelated  
systematic errors.
The systematic errors on \FLc take the correlations between
  the measurements into account using an off-set method: 
all correlated error sources, including the uncertainty
from the relative normalisation of the cross sections which 
in the extraction of \FLc is attributed to the $920$\,GeV 
cross sections, are considered separately 
and  added in quadrature to obtain the
total systematic error due to correlated sources. 
This error is added in quadrature
to the statistical and uncorrelated systematic uncertainties to obtain
the total error on $F_L$. 
The measurement is limited to bins where the total error 
is below $0.6$.

\begin{figure}[t]
   \centering
   \epsfig{file=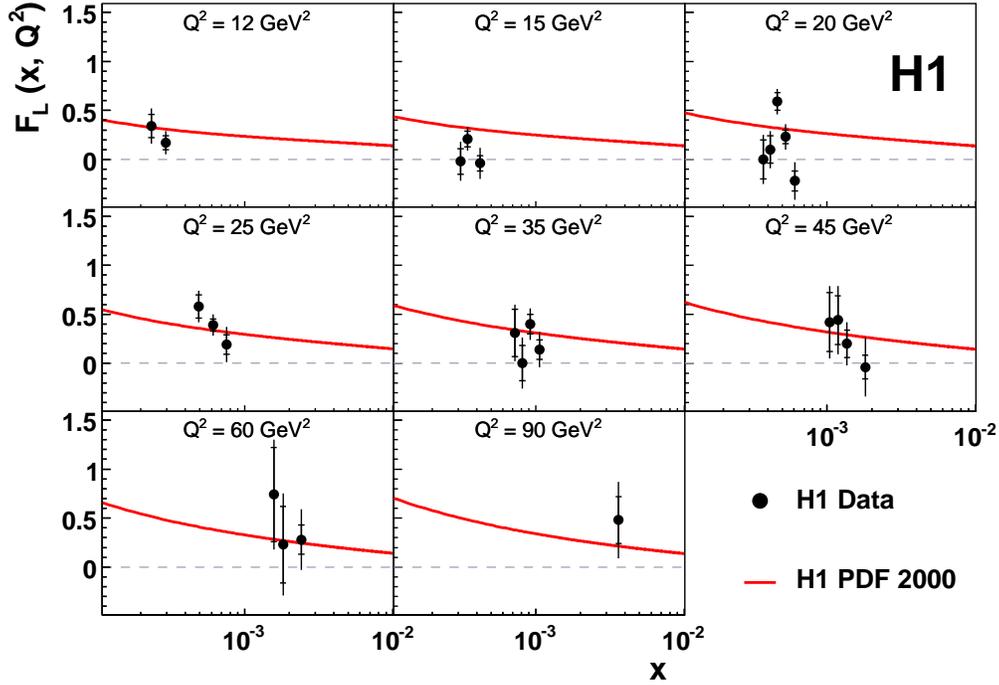,width=14cm} % 
   \caption{The longitudinal proton structure function $F_L(x,Q^2)$.
 The inner error bars denote the statistical error, the full
    error bars include the systematic errors. The luminosity uncertainty
    is not included in the error bars.  The curve represents
     the NLO QCD prediction derived from the  H1 PDF 2000 fit
      to previous H1 data.}
   \label{figfl}
\end{figure}

The measurement of $F_L(x,Q^2)$  is shown in figure\,\ref{figfl}.
The result is consistent with the prediction obtained with the
H1 PDF 2000 fit\,\cite{h1pdf2k}, which was performed using only the H1 high energy
cross section  data.  
The measurement is also consistent with previous
determinations of \FLc by H1\,\cite{alfs}, which used NLO QCD to 
describe and subtract the \Fc term from the measured
reduced cross section at high $y$.

\begin{figure}[h]
   \centering
   \epsfig{file=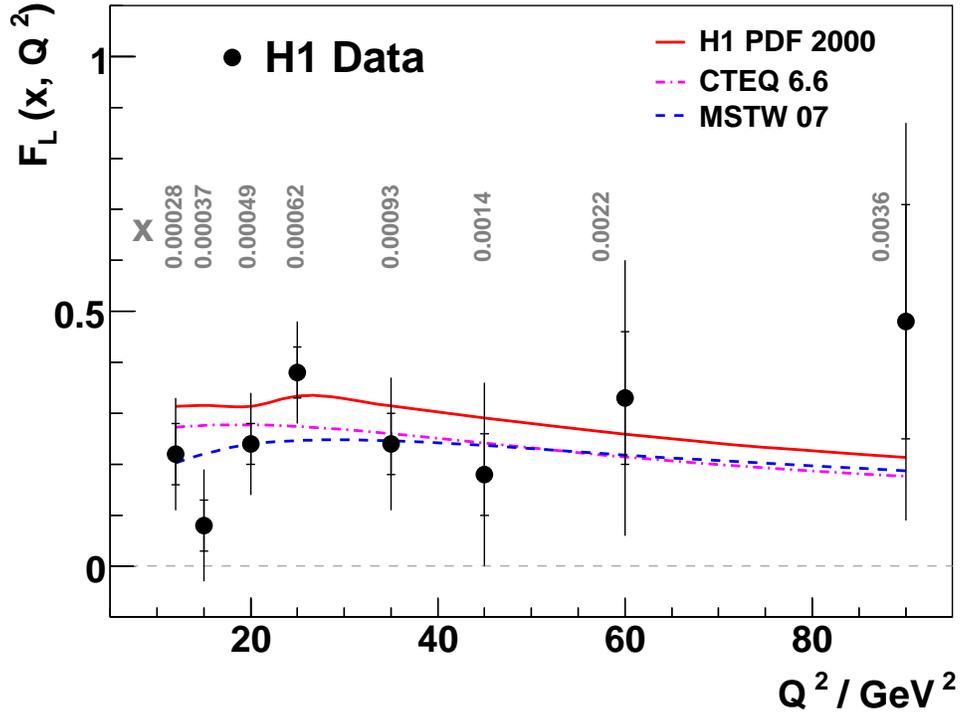,width=14cm} %
    \caption{The longitudinal proton structure function \FLc
    shown as a function of $Q^2$ at the given values of
     $x$. 
     The inner error bars denote the statistical error, the full
    error bars include the systematic errors. The luminosity uncertainty
    is not included in the error bars.
       The solid  curve describes the expectation
     on \FL  from the H1 PDF 2000 fit using NLO QCD. 
     The dashed (dashed-dotted) curve is
      the expectation of the MSTW (CTEQ) group using NNLO (NLO) QCD.
      The theory curves connect predictions at the 
      given $(x,Q^2)$ values by linear extrapolation.
}
   \label{figflav}
\end{figure}

The values on \FL resulting from averages over $x$ at fixed $Q^2$ 
are presented in  figure\,\ref{figflav}  and given in table\,\ref{tab:tabfl}.
The average is performed taking the $x$ dependent correlations 
between the systematic errors into account. The measurement of \FL is
compared with the H1 PDF 2000 fit and with
the expectations from global parton distribution
fits at higher order perturbation theory performed by the MSTW\,\cite{mrsw}  and the CTEQ group\,\cite{cteq} groups. Within the experimental
uncertainties the data are consistent with these predictions. 
This consistency underlines the applicability of the
DGLAP evolution framework of perturbative QCD at low Bjorken~$x$ at HERA.
% The measurement is therefore an important confirmation of the
% DGLAP evolution framework of perturbative QCD at low Bjorken~$x$ at HERA. 
%
\vspace{0.4cm}
\renewcommand{\arraystretch}{1.1}
\begin{table}[h]
\begin{center}
\begin{tabular}{|c|l|c|c|c|c|c|} \hline
 $Q^2$/GeV$^2$ & $\;\;\;\;\;x$ & $F_L$ & stat. & uncorr. & corr. & total \\
 \hline
         12  & 0.00028 &    0.22 &    0.06 &    0.05 &    0.08 &    0.11 \\
         15  & 0.00037 &    0.08 &    0.05 &    0.04 &    0.09 &    0.11 \\
         20  & 0.00049 &    0.24 &    0.04 &    0.04 &    0.09 &    0.10 \\
         25  & 0.00062 &    0.38 &    0.05 &    0.05 &    0.08 &    0.10 \\
         35  & 0.00093 &    0.24 &    0.06 &    0.06 &    0.09 &    0.13 \\
         45  & 0.0014   &    0.18 &    0.08 &    0.08 &    0.14 &    0.18 \\
         60  & 0.0022   &    0.33 &    0.13 &    0.13 &    0.19 &    0.27 \\ 
         90  & 0.0036   &    0.48 &    0.23 &    0.22 &    0.22 &    0.39 \\   \hline
\end{tabular}
\caption{The longitudinal proton structure function \FL measured at the given
values of $Q^2$ and $x$. The 
 statistical,  uncorrelated and  correlated systematic
uncertainties are given as well as the total uncertainty.}
\label{tab:tabfl}
\end{center}
\end{table}
\section{Summary}
This letter presents the first 
measurement of the longitudinal proton structure function 
in deep inelastic scattering
at low $x$.   The $F_L$ values
are extracted from three sets of cross section
measurements at fixed $x$ and $Q^2$, but different
inelasticity $y$, obtained with three different proton beam energies
at HERA. The results confirm DGLAP QCD predictions
for $F_L(x,Q^2)$,  determined from  previous HERA data,
which are dominated
by a large gluon density at low $x$.
At the current level of accuracy,  for the covered
$Q^2$ range between $12$ and $90$\,GeV$^2$, 
the data are thus consistent with perturbative QCD.
% a good  
%agreement is thus observed with perturbative QCD.

%\newpage
\section*{Acknowledgements}
We are grateful to the HERA machine group whose outstanding
efforts have made this experiment possible.
We thank the engineers and technicians for their work in constructing and
maintaining the H1 detector, our funding agencies for
financial support, the DESY technical staff for continual assistance
and the DESY directorate for support and for the
hospitality which they extend to the non DESY
members of the collaboration.


\begin{thebibliography}{99}
%
%\bibitem{isolprc05}
%H1 Collaboration,  Proposal to Switch to Positrons in January 2006, submitted to the PRC 11/2005.
%
\bibitem{cgross}
C.\,Callan and D.\,Gross, Phys. Rev. Lett. {\bf 22} (1969) 156.
%
\bibitem{am}
A.\,Zee, F.\,Wilczek and S.B.\,Treiman,  Phys. Rev. {\bf D10} (1974) 2881;\\
G.\,Altarelli and G.\,Martinelli, Phys. Lett. {\bf B76} (1978) 89.
%
\bibitem{dglap}
 V.N.\,Gribov and L.N.\,Lipatov,
 %``Deep Inelastic E P Scattering In Perturbation Theory,''
 Yad.\ Fiz.\  {\bf 15} (1972) 781
 [Sov.\ J.\ Nucl.\ Phys.\  {\bf 15} (1972) 438]; \\
  V.N.\,Gribov and L.N.\,Lipatov,
 %``E+ E- Pair Annihilation And Deep Inelastic E P Scattering In Perturbation Th
%eory,''
 Yad.\ Fiz.\  {\bf 15} (1972) 1218
 [Sov.\ J.\ Nucl.\ Phys.\  {\bf 15} (1972) 675]; \\
 Y.L.\,Dokshitzer,
 %``Calculation Of The Structure Functions For Deep Inelastic Scattering And E+ 
%E- Annihilation By Perturbation Theory In Quantum Chromodynamics. (In Russian),'
 Sov.\ Phys.\ JETP {\bf 46} (1977) 641
 [Zh.\ Eksp.\ Teor.\ Fiz.\  {\bf 73} (1977) 1216]; \\
 %
 G.\,Altarelli and G.\,Parisi,
 %``Asymptotic Freedom In Parton Language,''
 Nucl.\ Phys.\ {\bf B126} (1977) 298.
 %%CITATION = NUPHA,B126,298;%%
%\bibitem{dglap}
%dglap equations
%
\bibitem{rdata}
 J.J.\,Aubert {\it et al.}, EMC Collaboration, 
 Phys. Lett. {\bf B121} (1983) 87; \\
A.C.\,Benvenuti {\it et al.}, BCDMS Collaboration, 
  Phys. Lett. {\bf B223} (1989) 485; \\
L.W.\,Whitlow  {\it et al.}, Phys. Lett. {\bf B250} (1990) 193; \\
M.\,Arneodo {\it et al.}, NMC Collaboration,
 Nucl. Phys. {\bf B483} (1997) 3 [hep-ex/9610231].
%  CERN preprint CERN-EP/89-06.
%
%
\bibitem{fla}
 C.\,Adloff {\it et al.}, H1 Collaboration,  Phys. Lett. {\bf B393} (1997) 452 [hep-ex/9611017].
%
\bibitem{alfs}
 C.\,Adloff {\it et al.}, H1 Collaboration,  Eur. Phys. J. {\bf C21} (2001) 33 [hep-ex/0012053].
% [hep-ex/0012053].
%H1 Collaboration,  Paper contributed to ICHEP2004, H1prelim-03-043.
%
\bibitem{h1pdf2k}
C.\,Adloff {\it et al.}, H1 Collaboration,  Eur.\,Phys.\,J.\,{\bf C30} (2003) 1 [hep-ex/0304003].
%
\bibitem{mrsw}
A.D.\,Martin, W.J.\,Stirling, R.S.\,Thorne and G.\,Watt, 
%(University Coll. London) . IPPP-07-23, DCPT-07-46, Jun 2007. 13pp.
Phys. Lett. {\bf B652} (2007) 292 [hep-ph/0706.0459].
\bibitem{cteq}
J.\,Pumplin, H.L.\,Lai and  W.K.\,Tung,
Phys. Rev. {\bf D75} (2007) 054029 [hep-ph/0701220]; \\
P.M.\,Nadolsky {\it et al.},
%, H.-L. Lai, Q.-H. Cao, J. Huston, J. Pumplin, D. R. Stump, W.-K. Tung,
%C.-P. Yuan,
hep-ph/0802.0007.
%
\bibitem{willy}
      E.B.\,Zijlstra and W.\,van\,Neerven, Nucl. Phys. {\bf B383}
     (1992) 525; \\
      S.A.\,Larin and J.A.M.\,Vermaseren, Z.Phys. {\bf C57} (1993) 93.
\bibitem{jos}
S.\,Moch, J.A.M.\,Vermaseren and A.\,Vogt,
  %``The longitudinal structure function at the third order,''
  Phys. Lett. {\bf B606} (2005) 123, and references therein.
%
\bibitem{h1}
I.\,Abt  {\it et al.}, H1 Collaboration,  Nucl. Instr. Meth. 
{\bf A386} (1997) 310; \\
 I.\,Abt  {\it et al.}, H1 Collaboration, Nucl. Instr. Meth. 
 {\bf A386} (1997) 348.
%
\bibitem{cip}
J.\,Becker {\it et al.}, Nucl. Inst. Meth. {\bf A586} (2008) 190 [physics/0701002].
%
\bibitem{spacal}
R.\,Appuhn {\it et al.}, Nucl. Instr. and Meth. {\bf A386}
(1996) 397.
%
\bibitem{phojet}
R.\,Engel and J.\,Ranft, Phys. Rev. {\bf D54} (1996) 4244 [hep-ph/9509373].
%
\bibitem{django}
G.A.\,Schuler and H.\,Spiesberger, 
Proc. Workshop on HERA Physics, Vol 3, eds. W.\,Buchm\"uller and 
G.\,Ingelman, Hamburg, DESY (1992), p. 1419;\\
A.\,Kwiatkowski, H.\,Spiesberger  and H.-J.\,M\"ohring, Comp. Phys. Comm. {\bf 69} (1992) 155, \\
version 1.14 of DJANGOH is used.
%
\bibitem{ariadne}
L.\,L\"onnblad,  Comp. Phys. Comm. {\bf 71} (1992) 15, 
version 4.10 is used.
%
\bibitem{jetset}
T.\,Sj\"ostrand and M.\,Bengtsson,
%``The Lund Monte Carlo For Jet Fragmentation And E+ E- Physics. Jetset
%Version 6.3: An Update,''
Comput. Phys. Comm.\  {\bf 43} (1987) 367, 
%%CITATION = CPHCB,43,367;%%
version 7.4 is used.
%
\bibitem{geant}
R.\,Brun {\it et al.}, GEANT3, CERN Program Library, W5013.
%
\bibitem{HECTOR}
 A.\,Arbuzov {\it et al.},  Comput.  Phys.  Comm. {\bf 94} (1996) 128 [hep-ph/9511434].
\end{thebibliography}
\end{document}